\documentclass[runningheads]{svmult}

\usepackage{makeidx}   % allows index generation
\usepackage{graphicx}  % standard LaTeX graphics tool
\usepackage{subeqnar}  % subnumbers individual equations
\usepackage{multicol}  % used for the two-column index
\usepackage{physprbb}  % modified textarea for proceedings,
\makeindex             % used for the subject index
\newcommand{\greeksym}[1]{{\usefont{U}{psy}{m}{n}#1}}
\newcommand{\umu}{\mbox{\greeksym{m}}}

\begin{document}
\title*{The Phoenix Deep Survey: \protect\newline Evolution of Star Forming Galaxies}
\toctitle{The Phoenix Deep Survey:
\protect\newline Evolution of Star Forming Galaxies}
% allows explicit linebreak for the table of content
%
%
\titlerunning{The Phoenix Deep Survey}
% allows abbreviation of title, if the full title is too long
% to fit in the running head
%
\author{A.~M.~Hopkins\inst{1,2}
\and J.~Afonso\inst{3}
\and A.~Georgakakis\inst{4}
\and M.~Sullivan\inst{5}
\and B.~Mobasher\inst{6}
\and L.~E.~Cram\inst{7}}
\authorrunning{A. M. Hopkins et al.}
% if there are more than two authors,
% please abbreviate author list for running head
%
%
\institute{University of Pittsburgh, Department of Physics and Astronomy\\
3941 O'Hara St, Pittsburgh, PA 15206, USA
\and Hubble Fellow; ahopkins@phyast.pitt.edu
%\and University of Lisbon
%\and National Observatories of Athens
%\and University of Toronto
%\and Space Telescope Science Institute
%\and Australian Research Council
\and CAAUL, Observatory of Lisbon, Tapada da Ajuda, 1349-018 Lisbon, Portugal
\and National Athens Observatory, I.Metaxa \& Vas.Pavlou str., Athens 15236, Greece
\and University of Toronto, 60 St.\ George St, Toronto, Ontario M5S 3H8, Canada
\and STScI, 3700 San Martin Drive, Baltimore, MD 21218, USA
\and Australian Research Council, GPO Box 2702, Canberra, ACT 2601, Australia
}

\maketitle              % typesets the title of the contribution

\begin{abstract}
The Phoenix Deep Survey (PDS) is a multiwavelength survey based on deep
1.4\,GHz radio observations used to identify a large sample of star forming
galaxies to $z=1$. Photometric redshifts are estimated for
the optical counterparts to the radio-detected galaxies, and their
uncertainties quantified by comparison with spectroscopic redshift
measurements. The photometric redshift estimates and associated best-fitting
spectral energy distributions are used in a stacking analysis exploring
the mean radio properties of $U$-band selected galaxies. Average flux
densities of a few $\umu$Jy are measured.
\end{abstract}

\section{Introduction}
The study of galaxy evolution in recent years has included a strong
focus on the star formation properties of galaxies. Many of these studies are
based primarily on selection at ultraviolet (UV) and optical wavelengths,
known to be strongly affected by obscuration due to dust. It has
been shown that selection at these wavelengths results in samples of star
forming systems that miss a significant fraction of heavily obscured galaxies
\cite{2002ApJ...581..844S}. There have moreover been suggestions that the
most vigorous star forming (SF) systems suffer the most obscuration
\cite{2003astro.ph..7175A,2003astro.ph..6621H,2002AXA...383..801B,2001ApJ...558...72S,2001AJ....122..288H}.
Radio selection provides an efficient tool to construct a SF
galaxy sample free from dust induced biases, and the average obscuration in
such samples indeed appears significantly higher than in optically
selected samples \cite{2003astro.ph..7175A,2003astro.ph..6621H}.

Motivated to construct a homogeneously selected sample of SF galaxies,
unbiased by the effects of obscuration due to dust, the Phoenix Deep Survey
(PDS, see http://www.atnf.csiro.au/people/ahopkins/phoenix/) is based on
a deep ($60\,\umu$Jy), wide-area (4.5 square degree) 1.4\,GHz survey with the
Australia Telescope Compact Array.
This provides one of the largest existing deep 1.4\,GHz source catalogues
\cite{2003AJ....125..465H} containing a large fraction of SF
galaxies spanning the broad redshift range $0<z<1$. The PDS has already
been highly successful in providing a basis for several investigations
of the nature of SF galaxies and their evolution
(\cite{2003MNRAS.345..939G,2003astro.ph..7175A,2003AJ....125..465H} and
references therein). Throughout the present investigation
we assume a ($\Omega_M=0.3,\Omega_{\Lambda}=0.7,H_0=70$) cosmology.

\section{Photometric redshift analysis}

Deep $UBVRI$ observations of about one square degree within
the PDS have recently been analysed (these are described in detail in
\cite{2004astro.ph..XXXXS}). These data achieve a $5\sigma$ level of
$R_{\rm AB}\approx24.5$ and optical catalogues have been constructed and
cross-correlated with the 1.4\,GHz catalogues. This multicolour data has been
used to estimate photometric redshifts for all $\approx 40000$ galaxies
detected in each of the 5 bands. This includes about 800 optical counterparts
of the radio detected galaxies in this area.

For our analysis we use the photometric redshift code of Connolly et al.
\cite{1995AJ....110.2655C} (see also \cite{2000AJ....120.1588B}).
We use SED templates based on those of Coleman et al.
\cite{1980ApJS...43..393C} (hereafter CWW),
providing four standard SEDs (E/S0, Sbc, Scd, Im), extended in the UV
and IR wavelength regions using the GISSEL98 code \cite{1993ApJ...405..538B}.
Rather than using these four SEDs directly, we use the method of
optimal subspace filtering \cite{2000AJ....120.1588B} to provide a large
number (61) of smoothly interpolated SEDs based on the reference CWW SEDs.
This supports more realistic type estimates for most of the galaxies.
We allow the possible photometric redshifts to range from 0.0 to 1.3 and
also apply a prior constraining the absolute magnitudes of the
galaxies to the broad range $-29<M_B<-16$ (having the effect of removing
photometric redshift fits with unphysically high or low redshifts).

\begin{figure}[th]
\begin{center}
\centerline{\rotatebox{-90}{\includegraphics[width=9cm]{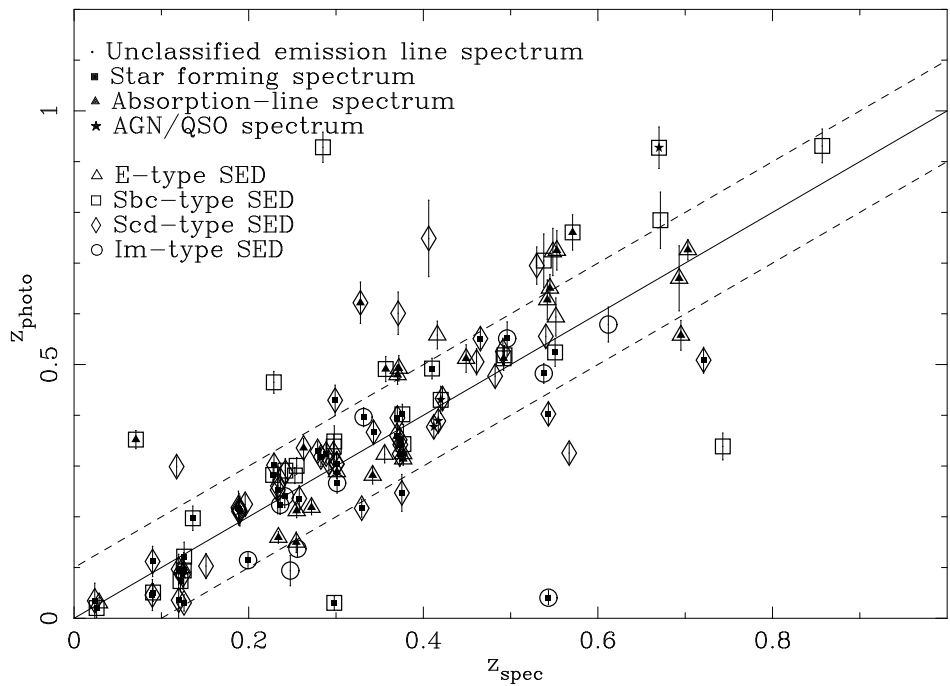}}}
\end{center}
\caption[]{A comparison of spectroscopic and photometric redshifts for the
spectroscopically observed sub-sample. The points and filled symbols refer
to the spectroscopic classification of the galaxies \cite{1999MNRAS.306..708G},
while the open symbols refer to the best-fitting SED from the photometric
redshift estimation. The open symbols are estimated by binning the 61
subspace filtered SEDs approximately into the closest CWW-type classification.
Note that this figure shows redshifts in linear units, with the dashed lines
indicating offsets of $\pm0.1$ in $z_{\rm photo}$ about the one-to-one line.}
\label{photoz}
\end{figure}

Figure~\ref{photoz} compares the photometric redshift estimate with
spectroscopic redshift for 116 radio sources with an optical counterpart
having both $UBVRI$ detections and spectroscopic data.
The filled symbols (including points) indicate
the spectroscopic classification, while the open symbols give an
estimate of the best-fitting SED template type (after binning the 61
subspace filtered templates into four bins, based approximately on
the closest CWW-type template). The reliability of the photometric
redshifts can be characterised in several ways. The rms of $|\Delta
z|=|z_{\rm photo}-z_{\rm spec}|$ is 0.1 for these 116 galaxies. The
rms of $|\log[(1+z_{\rm photo})/(1+z_{\rm spec})]|$ is 0.028
(implying a typical uncertainty of $7\%$ in $1+z_{\rm photo}$).
The rms of $|\Delta z|/(1+z_{\rm spec})$ is 0.065. This level
of reliability in the photometric redshifts compares favourably
with that of other analyses \cite{1999ApJ...513...34F,2003MNRAS.345..819R}.
As well as providing reasonable redshift estimates, the best-fitting SED is
also a good indicator of the galaxy type, in the sense that spectroscopic
absorption-line systems are mostly well-fit by early-type SEDs, while
SF systems are mostly well-fit by late-type SEDs.

\begin{figure}[th]
\begin{center}
\centerline{\rotatebox{-90}{\includegraphics[width=7cm]{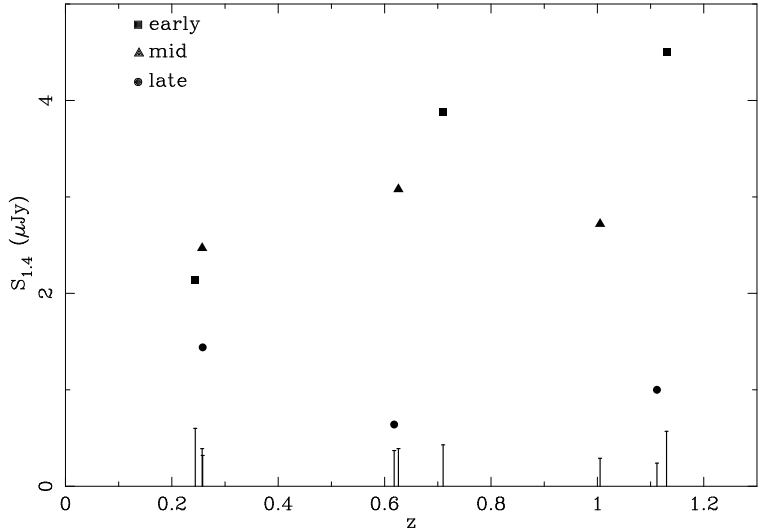}}}
\end{center}
\caption[]{Average 1.4\,GHz flux density inferred from the stacking
analysis for galaxies of early (squares), mid (triangles) and late (circles)
SED type. The flux densities are shown as a function of the median
photometric redshift of all the objects contributing to each of the nine
stacked images. The vertical bars below each point indicate the rms noise
level of the stacked image in which each measurement was made, with
detections ranging from $3.6-9\sigma$. The only non-detection is the
late-type SED (circle) in the middle redshift bin, and is shown to
indicate the upper limit for this stacked image (an arrow was omitted to
avoid clutter with the rms noise level bars).}
\label{svsz}
\end{figure}

\begin{figure}[th]
\begin{center}
\centerline{\rotatebox{-90}{\includegraphics[width=7cm]{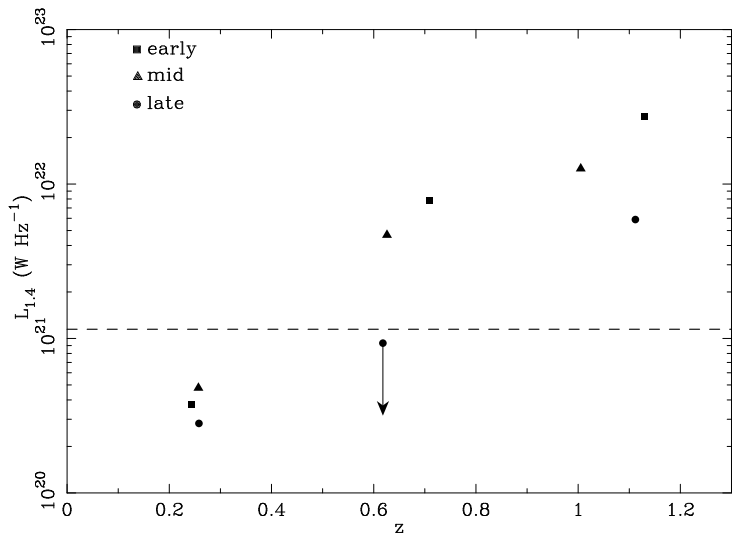}}}
\end{center}
\caption[]{Average 1.4\,GHz luminosity inferred from the flux density
measured in the stacking analysis, and the median photometric redshift. The
non-detection is now indicated as an upper limit. The dashed line
indicates the luminosity corresponding to an SFR of $1\,M_{\odot}\,$yr$^{-1}$
using the calibration of \cite{2003ApJ...586..794B}.}
\label{lvsz}
\end{figure}

\section{Stacking analysis of $U$-band galaxies}

The technique of stacking small subregions of an image at the
locations of a known population of objects that are not otherwise detected,
in order to extract a rough estimate of the mean emission properties of
a population, has been used with some success at X-ray wavelengths
\cite{2001AJ....122....1B,2002ApJ...576..625N}. This technique has been
applied to XMM observations of the PDS \cite{2003MNRAS.345..939G}
to explore the X-ray properties of radio-detected SF galaxies.
Following this success we extended the method to radio wavelengths,
performing a stacking analysis using the 1.4\,GHz mosaic image of the
PDS to explore the mean radio properties of extremely red galaxies (ERGs) not
otherwise detected at 1.4\,GHz \cite{2003astro.ph..9147H}. These results
have implications for the expected average radio luminosities and inferred
star formation rates of ERGs. Further investigation of the radio properties
of ERGs through stacking analyses are explored elsewhere in this volume
\cite{2003venice.conf...A}. We now further extend this technique to explore
the average 1.4\,GHz properties of a population of $U$-band selected galaxies,
which is expected to include a large fraction of SF galaxies. This will
provide an estimate of the typical radio properties for the population
of ``normal" or quiescent SF galaxies, as opposed to the starbursts
that often dominate studies of star formation in galaxies.

The available $U$-band data reaches a $5\sigma$ detection limit of
$U_{\rm AB}\approx25.0$. To explore radio flux density trends with both
redshift and galaxy type, we take advantage of the photometric redshifts and
best-fitting SED types to split the 40000 galaxies into three bins
in redshift and three bins in SED type (described as ``early," ``mid" and
``late," the CWW types Sbc and Scd being included in the ``mid" type).
The numbers of galaxies in each bin are given in Table~\ref{tab1}.
For the stacking analysis a subregion of $2'$ square was extracted
from the PDS 1.4\,GHz image at the location of each of the $U$-band galaxies.
To ensure that radio detections or uncatalogued low signal-to-noise (S/N)
radio emission do not bias the stacking signal, subregions are excluded
from the stacking analysis if the average 1.4\,GHz emission in a
$14''\times14''$ region centred at the location of the $U$-band galaxy is
above some S/N threshold. The threshold chosen was a fairly conservative
$1.5\sigma$, although the results change only marginally if slightly
higher thresholds ($2-3\sigma$) are used. Since the noise level is not
uniform across the PDS 1.4\,GHz image \cite{2003AJ....125..465H},
the individual subregions are weighted by the inverse square of the rms
noise background during the averaging step, in order to maximise the S/N
of the resulting stacked image.

The results of the stacking analysis are shown in Figures~\ref{svsz} and
\ref{lvsz}. Of the nine stacked images constructed, eight show
confident detections ($>3.6\sigma$). The exception is the middle redshift
bin for the late-type SED systems, where the peak flux at the expected
location of the source is $1.7\sigma$. The measured 1.4\,GHz flux densities
for the stacking results are shown in Figure~\ref{svsz} as a function
of the median photometric redshift for the objects contributing to
each final stacked image. The measured flux densities are of order
a few $\umu$Jy for each of the stacked detections, in rms backgrounds
around 0.3 to 0.5 $\umu$Jy. The ``mid"-type SED class seem to predominantly
show flux densities between the ``early" and ``late" types, consistent
with what might be expected if it was comprised of a combination of
both active galactic nuclei (AGNs) and SF systems, or of
systems driven by both processes.

The luminosities derived from these flux densities and
the median photometric redshifts are shown as a function of photometric
redshift in Figure~\ref{lvsz}. The trend to higher luminosities at higher
redshifts for all SED types is likely to be a result of our magnitude-limited
selection, since at high redshifts only the higher ($U$-band) luminosity
systems are being sampled, and the radio luminosity is likely to be correlated
at some level with the $U$-band luminosity \cite{2003astro.ph..6621H}.
The dashed line in this figure indicates a star formation rate (SFR) of
$1\,M_{\odot}\,$yr$^{-1}$ using the calibration of \cite{2003ApJ...586..794B}
(this line is about $40\%$ higher than if the calibration of
\cite{1992ARAXA..30..575C} were used). This shows that through the use
of the stacking technique it is possible to develop some insight into
the 1.4\,GHz properties of ``normal" or quiescent SF galaxies.
The fact that early type SED systems are detected at similar flux densities
may suggest either that low luminosity AGN are present in significant numbers
in normal galaxies, or that early type galaxies can support these low levels
of star formation, or even (since the flux density in early types seems
consistently higher than in late types) that both processes might be
occurring in these systems.

\begin{table}
\caption{Numbers of galaxies in redshift and type bins for the $U$-band stacking}
\begin{center}
\renewcommand{\arraystretch}{1.1}
\setlength\tabcolsep{5pt}
\begin{tabular}{cccc}
\hline\noalign{\smallskip}
Redshift range & Early & Mid & Late \\
\noalign{\smallskip}
\hline
\noalign{\smallskip}
 $0.00<z<0.43$ & 1358 & 3349 & 6377 \\
 $0.43<z<0.87$ & 3349 & 4009 & 3384 \\
 $0.87<z<1.30$ & 1562 & 6910 & 8663 \\
\hline
\end{tabular}
\end{center}
\label{tab1}
\end{table}

AMH acknowledges support provided by NASA through Hubble Fellowship grant
HST-HF-01140.01-A awarded by STScI.
JA acknowledges the support from the Science and
Technology Foundation (FCT, Portugal) through the fellowship
BPD-5535-2001 and the research grant POCTI-FNU-43805-2001.

%INDEX%%%%%%%%%%%%%%%%%%%%%%%%%%%%%%%%%%%%%%%%%%%%%%%%%%%%%%%%%%%%%%%
% Please check with the editor of your book whether he plans to
% include a "mutual" subject index - if so, please code your entries
% in the standard syntax. For your own purposes you may print your
% "personal" index by using the following commands:
%
%\clearpage
%\addcontentsline{toc}{section}{Index}
%\flushbottom
%\printindex
%%%%%%%%%%%%%%%%%%%%%%%%%%%%%%%%%%%%%%%%%%%%%%%%%%%%%%%%%%%%%%%%%%%%%

\end{document}